\documentclass[aps,superscriptaddress,twocolumn]{revtex4}%
\usepackage{amsfonts}
\usepackage{amsmath}
\usepackage{amssymb}
\usepackage{graphicx}
\usepackage{color}
\usepackage{float}
\begin{document}

\title{Amoeboid swimming in a channel}

\author{Hao Wu}
\affiliation{Universit{\'e} Grenoble Alpes, LIPHY, F-38000 Grenoble, France}
\affiliation{CNRS, LIPHY, F-38000 Grenoble, France}
\author{A. Farutin}
\affiliation{Universit{\'e} Grenoble Alpes, LIPHY, F-38000 Grenoble, France}
\affiliation{CNRS, LIPHY, F-38000 Grenoble, France}
\author{W.-F. Hu}
\affiliation{Department of Applied Mathematics, National Chung Hsing University, 145 Xingda Road, Taichung 402, Taiwan}
\author{M. Thi{\'e}baud}
\affiliation{Universit{\'e} Grenoble Alpes, LIPHY, F-38000 Grenoble, France}
\affiliation{CNRS, LIPHY, F-38000 Grenoble, France}
\author{S. Rafa\"i}
\affiliation{Universit{\'e} Grenoble Alpes, LIPHY, F-38000 Grenoble, France}
\affiliation{CNRS, LIPHY, F-38000 Grenoble, France}
\author{P. Peyla}
\affiliation{Universit{\'e} Grenoble Alpes, LIPHY, F-38000 Grenoble, France}
\affiliation{CNRS, LIPHY, F-38000 Grenoble, France}
\author{M.-C. Lai}
\affiliation{Department of Applied Mathematics, National Chiao Tung University, 1001 Ta Hsueh Road, Hsinchu 300, Taiwan}
\author{C. Misbah}
\affiliation{Universit{\'e} Grenoble Alpes, LIPHY, F-38000 Grenoble, France}
\affiliation{CNRS, LIPHY, F-38000 Grenoble, France}


\begin{abstract}
Several micro-organisms, such as bacteria, algae, or spermatozoa, use
flagellar or ciliary activity to swim in a fluid, while many other micro-organisms instead use ample shape
deformation, described as {\it amoeboid}, to propel themselves by either crawling on a substrate or swimming. Many eukaryotic cells  were believed to require
an underlying substratum to migrate ({crawl}) by using membrane deformation (like blebbing
or generation of lamellipodia) but there is now increasing evidence that a large variety of cells (including those
of the immune system) can migrate without the assistance of focal adhesion, allowing them to swim as efficiently as they can crawl.
This paper details the analysis of amoeboid swimming in a confined fluid by modeling
the swimmer as an inextensible  membrane deploying local active forces  (with zero total
force and torque). The swimmer {\color{black} displays} a rich behavior: it may settle into a
straight trajectory in the channel or navigate from one wall to the other depending on its confinement. The nature of the swimmer is also found to be affected by confinement: the swimmer can behave, {\color{black} on the average over one swimming cycle}, as a pusher
at low  confinement, and becomes a puller at {\color{black} higher} confinement, or vice versa. The swimmer's nature is thus not an intrinsic  property.
The scaling of the swimmer velocity $V$ with the force {\color{black} amplitude $A$} is analyzed in detail showing
that at small enough {\color{black} $A$, $V\sim A^2/\eta^2$} (where $\eta$ is the viscosity of
the ambient fluid), whereas  at large enough {\color{black} $A$}, $V$ is independent of the force and is determined solely by
the stroke cycle frequency and swimmer size. This finding starkly contrasts with currently known results found from swimming models where motion is based on ciliary and flagellar activity, where {\color{black} $V\sim A/\eta$}. To conclude, two definitions of efficiency as put forward in the literature are analyzed with distinct outcomes.
We find that one type of efficiency has an optimum as a function of confinement  while the other does not. Future perspectives are outlined.
\end{abstract}


\maketitle

\section{Introduction}

In the past decade, the problem of  swimming in a low Reynolds number environment
has attracted much attention from biology, physics, applied
mathematics, computer science, chemistry, and mechanical engineering. Many kinds of motile
unicellular micro-organisms living in the world of low Reynolds number \cite{Lauga2009,saintillan2012} use
a variety of strategies to propel themselves in liquids. So far most of the studies have been directed towards organisms employing flagellum and cilium activity. Some typical examples are
 spermatozoa~\cite{Fauci1995,Kantsler2013}, bacteria such as
{\it E. coli}~\cite{Lauga2006,Berke2008} and {\it Bacillus subtilis}~\cite{Sokolov2009} (examples of the so-called pusher), algae like {\it Chlamydomonas Reinhardtii} (an example of
puller)~\cite{Drescher2009,Guasto2010,Garcia2011,Kantsler2013}, {\it Volvox}~\cite{Drescher2009,Drescher2010}
and {\it Paramecium}~\cite{Jana2012,Zhang2015} (examples of neutral swimmers).
There has also been great interest in designing
 artificial microswimmers driven by different external fields, such as a rotating magnetic
field~\cite{Zhang2009}, a catalytic reaction field~\cite{Paxton2004}, and so on.

Another category of swimmers, much less studied until now, employ amoeboid swimming (AS).
Amoeboid movement  is the most common mode of locomotion in eukaryotic cells (monocytes,  neutrophils,  macrophages,   cancerous cells, etc.). The denomination {\it amoeboid motion} generically refers to motility due to shape deformation, and can be performed in a solution (like swimming) or on a substratum (crawling). {\it Eutreptiella Gymnastica}~\cite{Throndsen1969},
a common representative of euglenids of marine phytoplankton, is a prototypical example of an amoeboid swimmer. On the other hand, many eukaryotic  cells, such as leucocytes, fibroblasts, etc., which use ample membrane deformation for motility have been categorized as crawlers, {that is, as micro-organisms which use focal adhesion on a substratum to move forward.} Crawling
{\color{black} has been extensively studied} in the biological and biophysical literatures~{\color{black}\cite{Dambrosio2004,Entschladen2009}} for it plays an essential role in many physiological and pathological
processes including embryonic development, wound healing, immune response and cancer metastasis. {\color{black} It must be emphasized that there is yet no general consensus about the '{\it amoeboid}' denomination  for crawling cells. Some authors make a distinction between amoeboid cells and mesenchymal cells, where in the latter category the adhesion is so strong that the shape change may be minor and not appearing as a predominant factor for motion (see references \cite{Lammermann2009,Liu2015}). It has been seen that sometimes a cell may switch from mesenchymal to  amoeboid motion under confinement \cite{Liu2015}. In the present study amoeboid motion refers to any motion that requires a shape change to move forward.}

Several recent studies have been directed toward elucidating the role of focal adhesion in the migration process. One such study revealed that adhesion-based
crawling is not necessarily a pre-requisite for cell locomotion once leukocytes transmigrate across the
endothelium \cite{Lammermann2008}. Adhesion molecules would then serve only to stop moving leucocytes in the blood stream, making them adhere to the endothelium before transmigration. Studies {\it in vitro} have shown that integrins (the principal cell adhesion molecules) do not contribute to migration
in three-dimensional environments (in suspension with fibrous material, mimicking extracellular matrix of tissues)\cite{Lammermann2008}. {\color{black} Instead these cells migrate using actin polymerization, which leads to
membrane protrusion on the leading front.}
It is debated\cite{Liu2015,Hawkins2009} as to whether or not cell contact with fibrous material, without any specific
adhesion, is necessary in order to assist locomotion. A more recent study supports the idea of pure swimming
in cells of the immune system (neutrophils), and in dictyostelium  \cite{Barry2010};  it has been found that these cells, when far from the substrate, move {\color{black} toward a chemo-attractant. A remarkable result shows that they can swim \cite{Barry2010} with speeds of comparable magnitude to
those on a solid substrate (crawling)}. Other recent experimental studies focused \cite{Bergert2015} on cell
motion in a confined geometry showed that a blebbing cell\footnote{If the membrane becomes unstuck from the cortex, the intercellular pressure is able to push the membrane into a hemispherical blister, known as a bleb.} can undergo movement
through nonspecific adhesion, but only thanks to a {\color{black} solid-like friction from the substrate}. In other words,
this corresponds to a situation where the cell is in "physical" contact with the substrate on which
the cell might lean, in a fashion similar to the case of a cell leaning on a fibrous (or extracellular) material, but with no specific adhesion needed.

The above studies strongly support the idea that adhesion is not necessary for migration.
In some reported cases \cite{Barry2010} cells may even bypass contact with the substrate and perform pure swimming in a fluid. There seems thus to be
a whole spectrum of possibilities relevant for cell migration, ranging from specific-adhesion assisted
crawling, through pure non-specific friction with an elastic substratum, up to pure swimming in a solution. Thus an important task in research concerns bridging the gaps between pure swimming and motion by physical contact (solid-like friction), up to crawling.

Our goal is to first isolate different effects and then progressively increase the degree of complexity to reveal the imbricated nature of cell motility. To this end, we have systematically analyzed amoeboid motion in the pure swimming case and studied the effect of bounding walls.  Though there is no direct friction (no contact) with walls, the fluid between the swimmer and the walls serves as a support for transmitting frictional forces  between the swimmer and the walls. {\color{black} It must be noted, however, that while confinement (if not too strong in our case) promotes migration, the presence of bounding walls is not required; motility can take place in a completely unbounded medium, provided that the shape change is not reversible in time, as required by  the Purcell's scallop  theorem \cite{Purcell1977}.  }

Several models of directed self-propulsion by large shape deformation have been explored in the
literature~\cite{Avron2004,Ohta2009,Hiraiwa2011,Alouges2011,Arroyo2012,Vilfan2012,Loheac2013,Farutin2013,Wu2015}. In two recent studies we introduced a model in which \cite{Farutin2013,Wu2015}
the swimmer is considered as made of an inextensible  membrane enclosing a constant volume of liquid and exerting a set of normal forces leading to deformations of the swimmer and to propulsion. The first study \cite{Farutin2013} dealt with purely unconfined {\color{black} three dimensional (3D)}  geometry; the more recent work was dedicated to a confined swimmer in {\color{black} two dimensions (2D)} \cite{Wu2015}, a situation revealing several
puzzling phenomena which we will look at later. Here we report on several new issues emerging for a confined amoeboid swimmer. For example, we show that the velocity of migration has a nontrivial scaling with the force and address the
  notion of efficiency by testing two  definitions. {\color{black} We shall see, for example, that for certain confinement, the efficiency attains an optimal value.}
    We will discuss the space-scanning capability of the swimmer, where we find that generally the swimmer can scan about $90\%$ of the available space.

This paper is organized as follows: in Section \ref{intromodel}, we present a minimal model for AS.
In Section \ref{method} the two numerical methods used in this study are briefly detailed: the boundary integral method (BIM)~\cite{Thiebaud2013,Wu2015}
and the immersed boundary method (IBM)~\cite{Hu2014}.
In Section \ref{axialswim}, we present the main results regarding an axially moving swimmer. In Section \ref{Vnomono} we discuss both the peculiar behavior of the velocity as a function of confinement and the power consumption problem. Section \ref{effic} analyzes the swimmer efficiency. Section \ref{various} is devoted to  different dynamical motions. Conclusions and perspectives constitute the subject of Section \ref{conclusion}. Some technical details are given in an Appendix.

\section{Model and simulation method}
\label{intromodel}
\subsection{Amoeboid swimmer model}
\label{model}

\begin{figure}[t]
\begin{center}
\includegraphics[width=0.9\linewidth]{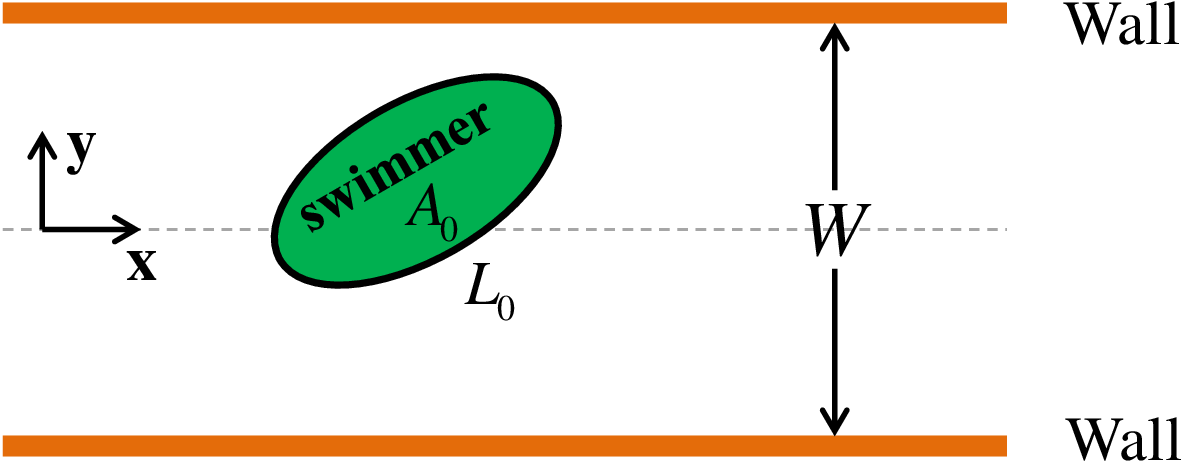}
\end{center}
\caption{\label{scheme}(Color online) A schematic representation of the system and some useful notations. The swimmer has internal area $A_0$ and perimeter $L_0$.}%
\end{figure}

For simplicity, and due to the high computational cost, we consider a 2D geometry. An amoeboid swimmer (AS) is
modeled as a membrane immersed in a Newtonian fluid domain of a given viscosity
$\eta$. The encapsulated fluid is taken to have the same viscosity,
for simplicity. We consider the pure Stokes regime (no inertia). An effective radius of the swimmer is defined as $R_0=\sqrt{A_0/\pi}$, where $A_0$ is
the enclosed area. We define {\color{black} an} excess perimeter (excess counted from a circular shape) $\Gamma=L_0/(2\pi R_0)-1$ to
describe the degree of deflation of swimmer shape.  $\Gamma=0$ represents a circular shape, whereas a large $\Gamma$ refers to a swimmer with ample shape deformations. The perimeter $L_0$ of the
swimmer membrane is taken to be constant due to the inextensibility of the membrane. Extensible membranes can be dealt with as well, but our goal is to consider a minimal model from which to capture basic features before refinement.
The swimmer is confined between two rigid walls separated by a distance $W$. We impose the no-slip boundary condition at
 the walls. A schematic representation of the studied system with some useful notations is shown in Fig. \ref{scheme}.


We consider certain forces deployed by the swimmer's internal machinery \footnote{The details of the internal machinery lay outside the scope of the present paper, but in the section devoted to discussion we shall comment on this point.},
with a certain distribution at the membrane. These are called the active forces and denoted as ${\mathbf F}_a$, and
depend on the position of a material point of the membrane and on time (see below). The active forces have a natural tendency to expand (or compress) the membrane which will {\color{black} resist} thanks to its intrinsic cohesive forces, resulting in a tension-like reaction. Since we are considering a purely incompressible membrane, the tension is such that
the local arclength should remain constant in the course of time.
\subsection{The choice of the active forces}
\label{activef}
The total force density (force per unit area) at the membrane is thus composed of an active part and a passive part, and can be written as
\begin{equation}
{\mathbf F}=F_a {\mathbf n} - \zeta c {\mathbf n} + \frac{\partial \zeta} { \partial s}{\mathbf t},
\label{Ftotal}
\end{equation}
\noindent{where ${ F_a} {\mathbf n}$ is the active force specified below and which we take to be directed along the normal direction, with ${\mathbf n}$ denoting
the unit normal vector.  $\zeta (s,t)$ is a Lagrange multiplier
that enforces local membrane inextensibility, $c$ is the curvature, ${\mathbf t}$ is the unit tangent vector, and $s$ is the arclength. {\color{black} Stresses (due to flow and active forces) vary, in general, from one membrane point to another, so
the membrane deploys a local opposing tension that tends to keep the local membrane length $ds$ constant in the course of time. In a vectorial form the tension is given by $\zeta (s) \mathbf t (s)$. Consider two neighboring membrane points at positions $s$ and $s+ds$, respectively. The resultant tension felt by the membrane element $ds$ is given by $\zeta (s+ds) \mathbf t (s+ds)- \zeta (s) {\mathbf t }(s)  =[(\partial \zeta /\partial s){\mathbf t} + \zeta \partial {\mathbf t} /\partial s ] ds + O(ds^2)$. The force per unit length is obtained by dividing the above result by $ds$. Using the geometrical result $\partial {\mathbf t} /\partial s = -c{\mathbf n}$ and taking the limit $ds\rightarrow 0$, we obtain the tension  force in eqn (\ref{Ftotal}).}

Since the boundary of the  swimmer is a closed curve, it is convenient to decompose the active force into Fourier series
\begin{equation}
{ F}_a(\alpha,t)= \sum _{k=-k_{max}, |k|\ne 0,1}^{k=k_{max}} \hat{F}_k(t) e^{ik\alpha},
\label{fourier}
\end{equation}
\noindent{where we have defined the normalized arc length $\alpha=2\pi s /L_0$.}
{\color{black} We must impose that the total force and the total torque equal zero:
\begin{equation}
\oint {\mathbf F} ds=0,\;\;\;  \oint {\mathbf r}\times {\mathbf F} ds=0.
\label{forcetorque}
\end{equation}
The integrals are performed over the swimmer perimeter.
This leads to three  equations  which are linear in terms of the active forces .
The coefficients of the system depend on the actual shape, which is unknown {\it a priori}. Conditions given by Eqn   (\ref{forcetorque}) impose linear relations between the amplitudes $\hat{F}_k(t)$. 
However, these relations did not allow us to easily reduce the number of independent force amplitudes because the resulting system was occasionally ill-conditioned. We solved this problem by adding an extra term to the active force, ${\mathbf F_0}+ \tilde F_0 \mathbf t.$ The first contribution is constant, and the second is purely tangential. We then use eqn (\ref{forcetorque}) to express $\tilde F_0 $ and ${\mathbf F_0}$ in terms of $\hat{F}_k's$. The virtue of this approach is the following: using the force and torque balance equations we easily see that in the linear system the coefficient in front of ${\mathbf F_0 }$ is proportional to the swimmer perimeter, whereas the coefficient in front of $\tilde  F_0 $ is proportional to the swimmer area, so that the matrix of the system of equations which multiplies the vector $({\mathbf F_0}, \tilde F_0 \mathbf t)$, is non-singular whatever the swimmer evolution, since the perimeter and area are definite positive quantities.}

Our strategy is to keep the model as simple as possible, and therefore we will limit the expansion (\ref{fourier}) to $k_{max}=3$.
The mode with $k=0$ does not play a role because the fluid is incompressible ($k=0$ corresponds to a homogeneous pressure jump across the membrane).
For simplicity we set  $F_1=0$.
We are thus left with
 two complex amplitudes $\hat{F}_2$ and $\hat{F}_3$ .
Once more for simplicity, the imaginary parts of $\hat{F}_2$ and $\hat{F}_3$ are set to zero, so the active force used in this
work is taken to be \footnote{Note that in Ref. \cite{Farutin2013} we also explored a large number of harmonics without affecting the essential results, lending support to the present simplifications.}
\begin{equation}
F_a(\alpha,t)= 2 F_2(t) \cos(2\alpha) + 2 F_3(t) \cos(3\alpha). \label{fact}
\end{equation}
$F_2$ and $F_3$ are time-dependent quantities expressing the cell's execution of a cyclic motion in order to move forwards. According to Purcell's theorem \cite{Purcell1977}, this cyclic motion should not be reversible in time owing to the invariance of the Stokes equations upon time reversal. Based on our first study on amoeboid motion \cite{Farutin2013} we shall make a simple choice for time dependence
 \begin{equation}
 F_2(t) =- A_2\cos(\omega t),\;\;\; F_3(t)= A_3\sin(\omega t).
 \end{equation}
 where for simplicity $A_2$ and $A_3$ are taken to be equal ($A_2=A_3=A$).  We are thus left with four parameters characterizing the force distribution,
$A$, the scalar   $\tilde F_0$ and the two components of ${\mathbf F_0}$. {\color{black} Use of condition (\ref{forcetorque})
allows one to express the three force components $ F_{0x}$, $F_{0y}$ and  $\tilde F_0$ as functions of $A$ for a given shape of the swimmer.
We are thus left with a single parameter monitoring the force, namely, the amplitude $A$.}



Note that the total force and the total torque of our passive force related to $\zeta$ (eqn (\ref{Ftotal})) vanish automatically.
Indeed, the passive force can be  written as a functional derivative of an energy
(see Ref.~\cite{Kaoui2008})
\begin{equation}
{\mathbf F}^{\rm mem}= -\frac{\delta E({\mathbf r}(s))}{\delta {\mathbf r}},
\end{equation}
where $\delta/ \delta{\mathbf r}$ is the functional derivative. Since this energy is intrinsic to the membrane, any rigid translation or rotation leaves it invariant. Let $\delta {\mathbf r}$ be the displacement of a given point on the membrane during a time interval $\delta t$, corresponding to a rigid translation or rotation. In both cases the energy $E({\mathbf r}(s))$ is unchanged and we have

\begin{equation}
    \delta E({\mathbf r})=E({\mathbf r}+\delta {\mathbf r})-E({\mathbf r})=\oint \left[{\mathbf F}^{\rm mem}(s)\cdot \delta{\mathbf r}(s)\right] d s =0. \label{ft}
\end{equation}
For a translation, $\delta {\mathbf r}$ is constant and is denoted as $\delta {\mathbf r}_0$, whereas for a rotation we have $\delta {\mathbf r}=\boldsymbol{\omega}_0 \delta t \times {\mathbf r}\equiv \delta\boldsymbol{\theta}_0\times {\mathbf r}$, where ${\omega}_0$ is the angular velocity. Applying condition (\ref{ft}) to both cases one obtains

\begin{widetext}
\begin{equation}
\delta E({\mathbf r})=\left\{
\begin{array}{ll}
 \delta{\mathbf r}_0\cdot\oint_{\partial \Omega} {\mathbf F}^{\rm mem}(s) d s=0 & (\delta{\mathbf r}=\delta{\mathbf r}_0),
 \\
 \delta\boldsymbol{\theta}_0\cdot\oint_{\partial \Omega} \left[{\mathbf r}(s)\times{\mathbf F}^{\rm mem}(s)\right] d s=0 & (\delta{\mathbf r}=\delta\boldsymbol{\theta}_0\times {\mathbf r}).
\end{array}
\right. \label{force_torque}
\end{equation}
\end{widetext}


An explicit  proof of the fact that the passive forces do not contribute to the total force is given in the Appendix.

\subsection{Independent dimensionless parameters}

\quad To quantify confined amoeboid dynamics, we define a set of dimensionless parameters. The set of physical parameters  are the swimmer area $A_0$ and perimeter $L_0$ (recall that the area  defines a length scale $R_0=\sqrt{A_0/\pi}$), the channel width $W$, the viscosity of the fluid (taken to be the same outside and inside the swimmer) $\eta$, {\color{black} and two time scales $T_c$ and $T_s$. $T_c=\eta/A$  reflects the time needed for the swimmer to adapt to a static distribution of active forces of characteristic amplitude $A$.
 The larger the active force, the faster the shape adapts.} $T_s= 2\pi /\omega$ is the stroke period. All together, we obtain  a set of three dimensionless parameters ($\Gamma$  was already introduced above, but is regrouped here with other dimensionless numbers for the reader's convenience)
\begin{equation}
  \begin{cases}
    C={2R_0\over W},\\
    S={T_{s}\over 2\pi T_{c}}={A\over {\omega \eta}},
    \\
 \Gamma={L_0\over 2\pi R_0}-1.
 \label{dimension}
  \end{cases}
\end{equation}
The first parameter defines confinement strength, the second measures the ratio between the stroke period { and the adaption time}. A large $S$ means that the stroke is so slow that the shape has ample time to adapt to the applied force,
whereas a small $S$ means that the stroke is so fast that the shape does not have enough time to fully adapt to the applied force. Finally, recall that the third parameter is the excess length; large $\Gamma$ indicates a very deflated swimmer.

\section{Numerical methods}
\label{method}
We make use here of two methods: the boundary integral formulation~\cite{Thiebaud2013}
associated with the Stokes flow, and the immersed boundary method~\cite{Hu2014}, which  incorporates inertial effects, as well as arbitrary shape
of the boundaries and their compliance, for future studies.  We shall briefly recall below the two methods.


\subsection{Boundary integral method with two parallel plane walls}
\quad The general discussion  of the BIM can be found in~\cite{Pozrikidis1992} and application to the present problem in \cite{Thiebaud2013}. Due to the linearity
of the Stokes equations we can convert the bulk equations into a boundary integral equation which relates the velocity on the membrane to the forces exerted on the membrane
\begin{equation}
{\color{black} {\mathbf u} ({\mathbf r}_m,t)={1\over 4\pi \eta} \oint {\mathsf G} (\mathbf{r}^{\prime}_m,{\mathbf r}_m){\mathbf F} ({\mathbf r}^{\prime}_m,t) {\rm d} s({\mathbf r}^{\prime}_m)},
\label{greenW}
\end{equation}
where the integral is performed over the membrane, with ${\mathbf r}_m$ the membrane instantaneous position, $ {\mathbf u} ({\mathbf r}_m,t)$ is
the instantaneous velocity of a membrane point, ${\mathbf F}$ is the total force from  the membrane (eqn (\ref{Ftotal})). The unknown Lagrange multiplier $\zeta$ entering the total force (eqn (\ref{Ftotal})) is determined by requiring the divergence of velocity field along the membrane to be zero (membrane incompressibility condition).
{\color{black} ${\mathsf G}(\mathbf{r}^{\prime}_m,{\mathbf r}_m)$}
is the Green's function which vanishes at the bounding walls (i.e. at $y=\pm W/2$) \cite{Thiebaud2013}. In  an unbounded domain  the Green's function is nothing but the Oseen tensor,
which reads in 2D
\begin{equation}
\label{green_freespace}
 G_{ij}(\mathbf{x},\mathbf{y}) = -\delta_{ij} \ln r + \frac{r_i r_j}{r^2} ,
\end{equation}
where $r_i={x}_i-{y}_i$, and $r=r_i^2$ (Einstein convention is adopted).
However, this function does not vanish at the wall. Therefore two alternatives are possible: (i) either use this function,
and in which case we have to add to (\ref{greenW}) an extra integral term (the integrand contains  the product of the Green
function and the hydrodynamic force on the boundaries), see Ref. \cite{Cantat2003}, or (ii) find a Green function that vanishes at
the walls.  No explicit Green's function vanishing at the walls is available, but it can be  expressed in terms of a Fourier integral~\cite{Thiebaud2013}.
{\color{black} The advantage of the latter alternative is the absence of integral terms on the walls (the integral is performed on the swimmer membrane only),
and therefore the system size along the wall ($x$ direction) can be taken as infinite, thus avoiding any potential numerical problem
(like dependence of the results on system size) related to boundary condition along $x$.}
Some of the technical details of how the problem is numerically solved can be found in Ref. \cite{Thiebaud2013}.

\subsection{Immersed boundary method}

We present now the other method which can allow, in principle, to have arbitrary bounding geometry as well as deformable walls, a work which is planned in the future. Therefore, it is essential to compare the two methods in this paper, before exploring the versatility of the immersed boundary method.
\quad The immersed boundary formulation is an Eulerian-Lagrangian framework in which fluid
variables are defined in Eulerian manner, while the membrane related variables are defined in Lagrangian
manner. The swimmer membrane as a moving inextensible interface, immersed in a
$2$D fluid domain $\Omega$, is represented in a Lagrangian parametric form as
$\mathbf{R}(\alpha,t)=(X(\alpha,t),Y(\alpha,t))$, where $\alpha\in[0,2\pi]$
is a parameter of the initial configuration of the membrane. Thus,
the governing equations of the Navier-Stokes fluid system are

\begin{widetext}
\begin{equation}
\left\{
\begin{array}{ll}
\rho\left(\frac{\partial \mathbf{u} }{\partial t}+(\mathbf{u}\cdot\nabla)\mathbf{u}\right) = -\nabla p + \eta\Delta\mathbf{u} + \int_0^{2\pi}\mathbf{F}(\alpha,t)\delta(\mathbf{r}-\mathbf{R}(\alpha,t))|\mathbf{R}_{\alpha}|\rm{d}\alpha & \mbox{in } \Omega,
\\
\nabla\cdot\mathbf{u} = 0 &\mbox{in } \Omega,
\\
\frac{\partial\mathbf{X}(\alpha,t)}{\partial t} = \mathbf{U}(\alpha,t) = \int\mathbf{u}(\mathbf{r},t) \delta(\mathbf{r}-\mathbf{R}(\alpha,t))\rm{d}\mathbf{r} &\mbox{on } {\partial \Omega}.
\end{array}
\right. \label{imme}
\end{equation}
\end{widetext}

\noindent{where $\rho$ represents the mass density, and the Eulerian fluid and Lagrangian immersed variables are linked
by the $2$D Dirac delta function $\delta(\mathbf{r})=\delta(x)\delta(y)$.} The total membrane force $\mathbf{F}$
consists of the same active force as in BIM and the tension force. 
{\color{black} For this numerical method,  we adopt the spring-like elastic force to
approximate the membrane inextensibility, that is, taking $\zeta=\xi_0(ds/d\alpha -L_0/(2\pi))$ with a large stiffness constant $\xi_0$. In other words, the spring relaxes to its equilibrium length on  a very short time scale as compared to any physical time scale (e.g. shape deformation time scale)}. Details of numerical procedure can be found in references \cite{Hu2014,Wu2015}.

We perform the time-stepping scheme as follows: The governing
equations are discretized by the IBM. We consider the
computational domain as a rectangle $\Omega=[a,b]\times[c,d]$; the
no-slip boundary condition for the velocity is imposed on the two
walls $y=c,d$ and no-flux condition is applied on $x=a,b$. Within this domain, a uniform Cartesian
grid with mesh width $h$ in both $\mathbf{x}$ and $\mathbf{y}$ directions is employed. {\color{black} All simulations are performed by using an aspect ratio (length of the channel divided by the width) of a value 8. Comparison with BIM method -- with a literally infinite extent along $x$ -- shows good agreement. Smaller values of the aspect ratios lead to deviations from BIM where the swimming velocity is found to be higher, whereas the qualitative features remain unaffected.}
The fluid variables are defined on the standard staggered
marker-and-cell (MAC) manner~\cite{Harlow1965}. That is, the velocity
components $u_x$ and $u_y$ are defined at the cell-normal edges
$(x_{i-1/2},y_j) = (a+(i-1)h,c+(j-1/2)h)$ and $(x_i,y_{j-1/2}) =
(a+(i-1/2)h,c+(j-1)h)$ respectively, while the pressure $p$ is
defined at the cell center $(x_i,y_j) = (a+(i-1/2)h, c+(j-1/2)h)$.
For the membrane interface, we use
$\alpha_k=k \Delta \alpha, k=0, 1,\ldots, M$ with $\Delta \alpha = 2 \pi/M$ to
represent the Lagrangian markers $\mathbf{R}_k=\mathbf{R}(s_k)$ so that any
spatial derivatives can be performed spectrally accurate by using
Fast Fourier Transform (FFT) \cite{Trefethen2000}. 
Detailed numerical algorithm can be found in a previous work~\cite{Hu2014}.

\section{Axially moving swimmer}
\label{axialswim}

We first consider an amoeboid swimmer (AS) initially located in the center
of the channel, that is, its initial
center of mass is $(X_c,Y_c)=(0,0)$. Although this axial motion is generally found to be an unstable position (see {\color{black} Section \ref{various}}),
it gives useful insights into the dynamics of the confined amoeboid swimmer.  Figure~\ref{fig1} shows a typical snapshot over $170$
deformation cycles. For quite a long  period of time (in our simulation this solution survives for a few hundred cycles),
the swimmer moves along the channel ($x$ direction).
\begin{figure}[b]
\begin{center}
\includegraphics[width=0.9\linewidth]{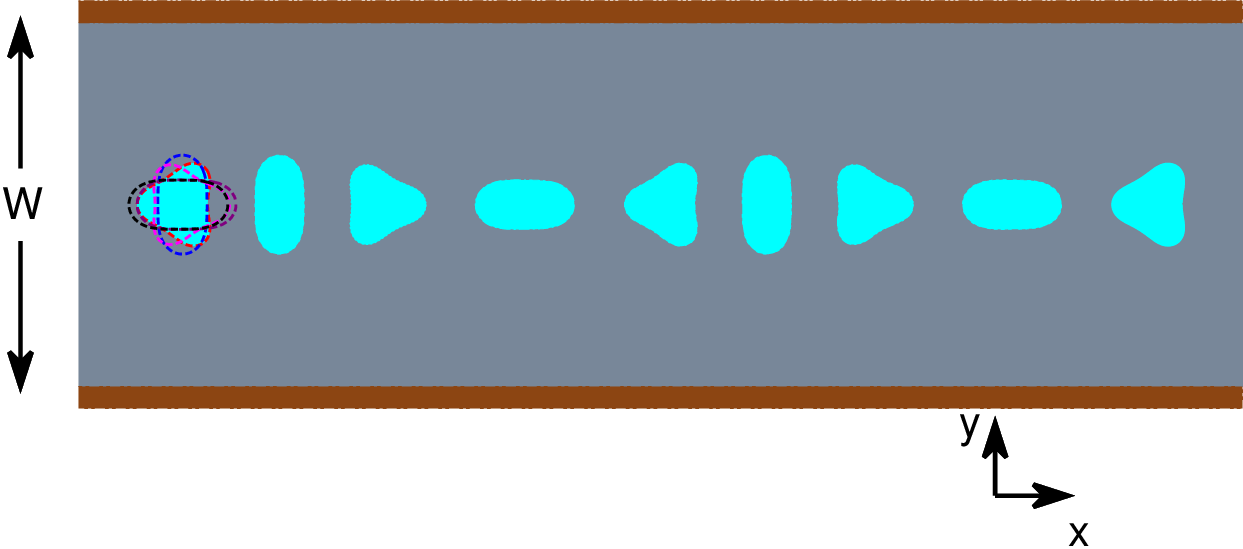}
\end{center}
\caption{\label{fig1}(Color online) Snapshots of an axially moving swimmer over time ($W=10.0R_{0}$).
The dashed profiles show a complete period $T_s$ of deformation and then a few shapes are represented
over a time of the order of $170 T_s$. $\Gamma=0.085$, $S=10$. }%
\end{figure}

In physical units the  average velocity of the swimmer will be defined as $\langle V \rangle= [X_c(t+T_s)- X_c(t)]/T_s$ (this is the net displacement of the center of mass over one cycle divided by the stroke cycle period).
We will define the dimensionless instantaneous velocity as $\bar{V}=VT_s /R_0$. Its average over one swimming cycle is defined as $\langle\bar V\rangle= \bar X_c(t+T_s)-\bar X_c(t)$, where $\bar X_c=X_c/R_0$.

Figure~\ref{fig2} shows  the center of mass of the swimmer, $\bar X_c(t)$,
undergoing a periodic oscillation in the axial direction  as shown in  (a).
A first observation is that for $C$ not too large (from $C=0.01$ up to $C=0.71$), the amplitude of the
oscillation increases with confinement. The slope of the curves $\bar X_c(t)$ also increase with $C$,  indicating an enhancement of migration speed
upon increased confinement. The increase of oscillation amplitude with $C$ (for not too large $C$) is already an interesting indication that the swimmer takes advantage of the walls to enhance its migration speed.

\begin{figure}[ptb]
\begin{center}
\includegraphics[width=0.9\linewidth]{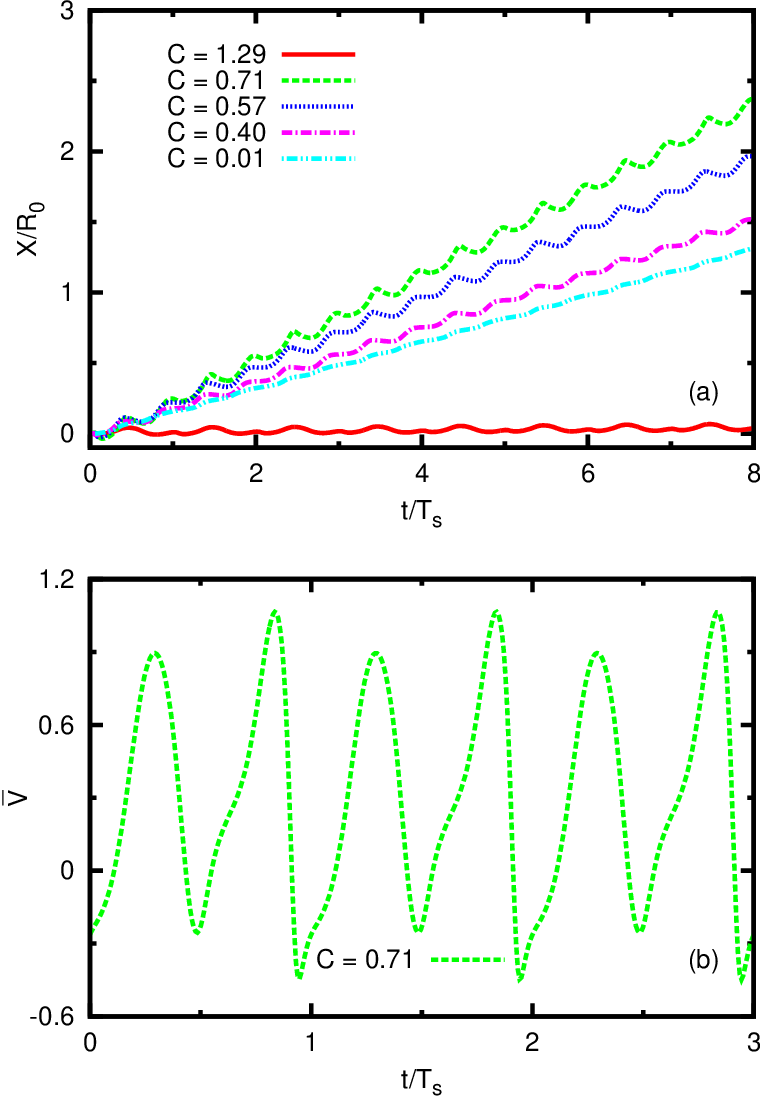}
\end{center}
\caption{\label{fig2} (Color online) (a) An axially moving amoeboid swimmer
undergoes a periodic oscillation in the axial direction ($x$ direction) of the channel as a function of time.
From strong to weak confinements, the corresponding oscillatory motions are represented by
different types of colorful lines. (b) Velocity of an axially moving swimmer as a function of time. $\Gamma=0.085$, $S=10$.}%
\end{figure}

\begin{figure}[ptb]
\begin{center}
\includegraphics[width=0.9\linewidth]{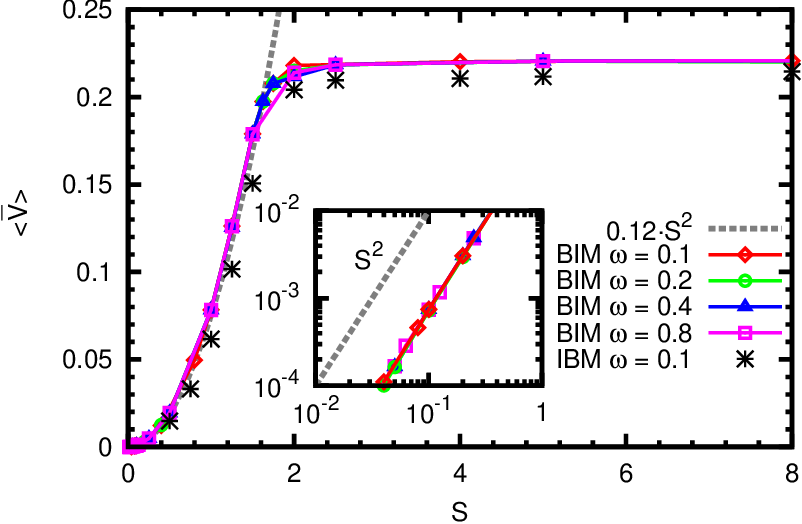}
\end{center}
\caption{\label{fig8} {(Color online) (a) The time-averaged velocity magnitude attains
a plateau with the increasing dimensionless amplitude of active forces $S$; $\Gamma=0.085$ and $C=0.5$.
 The inset shows the scaling  $S^2$ (grey dashed line) in a $\log-\log$ plot. The stars are data obtained by IBM, the other data are obtained by BIM.} }%
\end{figure}

\begin{figure}[ptb]
\begin{center}
\includegraphics[width=0.9\linewidth]{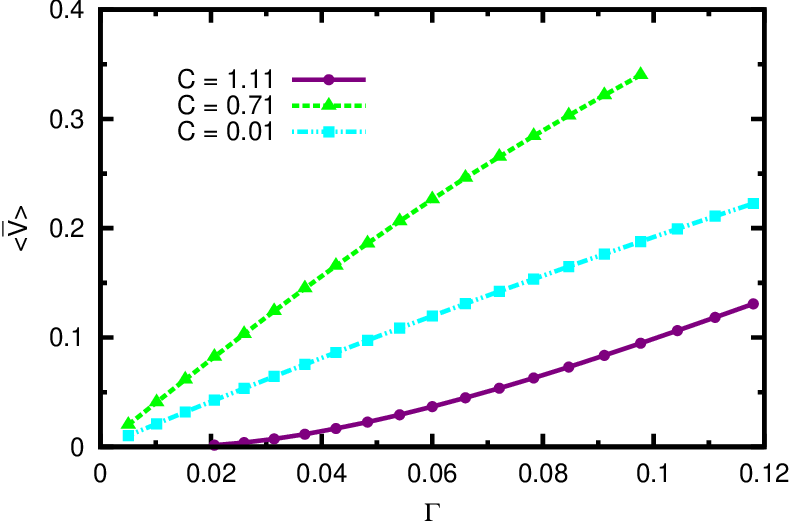}
\end{center}
\caption{\label{fig4} (Color online) In different confined regions, the linear and
nonlinear relations between the excess length $\Gamma$ and the velocity magnitude $\bar{V}$ are
calculated; $S=10$.}%
\end{figure}

\begin{figure}[ptb]
\begin{center}
\includegraphics[width=0.9\linewidth]{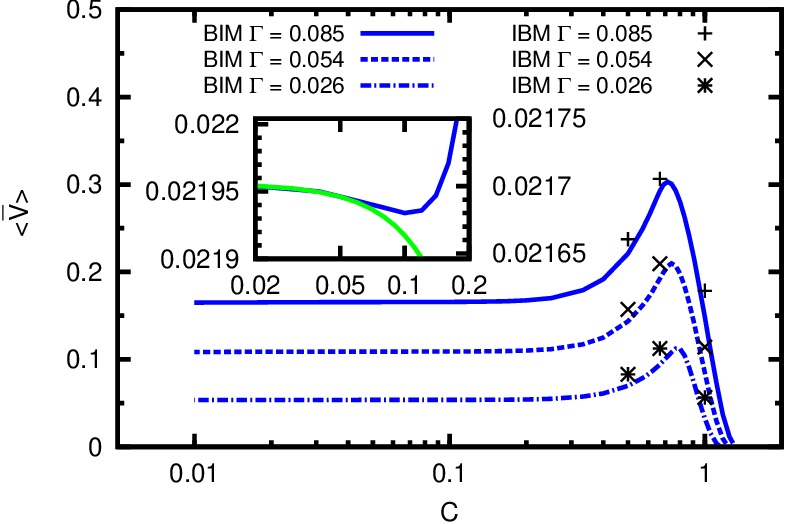}
\end{center}
\caption{\label{fig3} (Color online) Time-averaged velocity magnitudes as a function of $C$
in $\log$ scale for different $\Gamma$s.
The data represented by the solid lines are calculated by BIM, while the data represented by symbols are calculated by IBM; $S=10$. The inset shows a zoom for weak confinement with $\Gamma=0.010$ showing a weakly decreasing velocity (scale on the left side) and its comparison with the analytical theory in green (see also section \ref{Vnomono}), with the scale on the right side. The full numerical analysis and the analytical result agree quite well as long as the confinement is weak enough, in line with the assumption used to find the analytical solution.}
\end{figure}
Fig.~\ref{fig2} (b) shows that the instantaneous velocity changes
during three deformation stroke periods. The negative velocity is induced by the confining walls;
in the unconfined case the velocity always remains
positive~\cite{Farutin2013}.

\subsection{Scaling of the swimming velocity with the force amplitude}

Due to the linearity of the Stokes equations, one may expect that the velocity of the swimmer scales linearly with the force.
This has been adopted as a hypothesis in several studies where the  swimmer is {\color{black} modeled as rigid body}. Typical examples are the medeling of {\it E. coli}~\cite{Lauga2006,Berke2008}, {\it Bacillus subtilis}~\cite{Sokolov2009} (examples of the so-called pushers), and {\it Chlamydomonas} (an example of puller)~\cite{Drescher2009,Guasto2010,Kantsler2013}. In the present study  the relation between the velocity and the force is extracted a posteriori.  We find that  the velocity is a nonlinear function of the force amplitude. We have defined in equation (\ref{dimension})
a set of three dimensionless parameters. Our goal in this section being to discuss the effect of the force amplitude, we
set $\Gamma$ and $C$ to constant values. The physical (i.e. not the dimensionless) average velocity over a swimming cycle is given, from dimensional analysis, by the relation
\begin{equation}
\langle V\rangle \sim {R_0\over T_s} G(S).
\label{general}
\end{equation}
where $G(S)$ is a scaling function, unknown for the moment.  Let us first provide some bases for the behavior of $G(S)$ using physical intuition. An obvious guess is that for a large $S$, $\langle V\rangle$
should be independent of the force amplitude. Indeed, recall that $S$ can be  written as  $S\sim  T_s/T_c$. This is the ratio between the stroke period and the shape adaptation time. A large $S$ (i.e. a large force)   implies  that the swimmer attains its saturated shape (in response to applied forces) in a shorter time than the stroke time interval. In other words, further increasing the force amplitude will hardly change the sequence of shapes explored by the swimmer  during one stroke cycle.
 This increase of $A$ does not promote faster swimming, only a faster adaptation time. Indeed, in the Stokes regime time does not matter - only the configuration is important. Therefore, the speed should attain a plateau at large $S$. Our numerical simulation clearly shows this behavior (Fig.~\ref{fig8}). We can thus conclude that for large $S$, $\langle V\rangle\sim \omega R_0$. The velocity is thus solely determined by the stroke frequency.

The situation is less obvious for small $S$, the limit in which the force configuration changes in time faster than the shape adaptation. Increasing $S$ (but remaining in the small range limit) means the shape has more time to adapt (though not yet fully), resulting in a faster and faster motion since the shape, for each increase of $S$, will be {deformed} more and more until saturation (when the force is large enough).
The first natural expectation would have been that we would see a linear relationship between velocity and force, but this naive expectation does not comply with the numerical finding.
Our results show a quadratic behavior. A qualitative argument in favor of this behavior is the following:  {\color{black} if $A$ is small, it is legitimate to perform a perturbation theory in powers of $A$. The swimmer deformation (due to active force $F_a(\alpha, t)$ given by equation (\ref{fact})) is proportional to leading order to $F_a$, as is the swimmer velocity. Averaging the velocity over a cycle gives zero, since $F_a(\alpha, t)$ is proportional to a superposition of $\cos(\omega t)$ and $\sin(\omega t)$. Therefore the first non-vanishing contribution
is quadratic in $F_a$, leading to a quadratic dependence on $A$. Quadratic behavior was also recently reported for the  three-bead swimmer \cite{Pande2015}. However, no  saturation regime was shown in that paper.}

In summary, the velocity of the swimmer behaves as
\begin{equation}
\langle V\rangle \sim\left\{
\begin{array}{ll}
  {(R_0/T_s)} S^2 = {R_0T_sA^2 /\eta^2} & (S \ll 1),
 \\
 { R_0 / T_s} & (S \gg 1).
 \label{scalinsmallf}
\end{array}
\right.
\end{equation}
 Figure \ref{fig8} shows the full nonlinear behavior obtained from the numerical solution which {\color{black} is} in agreement with the above scalings. Note that the velocity behaves
for small force as $1/\eta^2$, which is distinct from the classical Stokes result where the velocity is $\sim 1/\eta$. This behavior is amenable to experimental testability (provided that the swimmer operates at a fixed given force when viscosity is varied). It is not however clear if the condition of $S\ll 1$ is abundant or not in the amoeboid
swimming world.
It is  thus   of great importance to  conceive of artificial amoeboid swimmers that {\color{black} could} be
monitored in order to scan the whole range of $S$ so as to permit tests of the above scaling relations.

\subsection{Velocity of the swimmer as a function of confinement and excess length}
We have  determined the velocity as a function of the deflation of the swimmer, a deflation characterized by the excess length $\Gamma$.
Quasi-linear and nonlinear relations between the center of mass' velocity
magnitude $\bar{V}$ and the excess length $\Gamma$ are found depending on confinement, as shown
in Fig.~\ref{fig4}.
We expect that the more the swimmer is deflated the larger its velocity. Indeed, the deformation amplitude of the swimmer increases with $\Gamma$, which thus implies a higher speed.
From Fig.~\ref{fig4}, it is found that the exponent
 for weak confinement is almost $1$ (shown by a cyan
line in Fig.~\ref{fig4}), and agrees with the study for a completely unbounded swimmer~\cite{Farutin2013}.
However, as confinement increases the picture is more complex: we have, at small $\Gamma$, a power law with exponent lower than $1$  (green line) {\color{black}   in the intermediate confinement regime,
and an exponent higher than 1 in the case of strong confinement (shown by a purple line in Fig.~\ref{fig4})}.

A prominent property we have recently reported on \cite{Wu2015} is the velocity's non-monotonous behavior
with confinement $C$. The results are summarized in Fig.~\ref{fig3}, where we compare the outcome  of the two numerical methods showing good agreement. Both methods show a maximum in the $\langle \bar V(C)\rangle$ curve.

\section{On the non monotonous behavior of the swimmer velocity as a function of $C$}
\label{Vnomono}
Once we have clarified the role of the force amplitude in the swimming speed, we would
like to discuss the behavior of the velocity as a function of confinement (Fig.\ref{fig3}) .
We will first provide a brief summary of wall effects reported so far in the literature before discussing our results.
The fact that the wall enhances motility is reported in several papers \cite{Felderhof2010,Jana2012,Zhu2013,Bilbao2013,Ledesma2013,Acemoglu2014,Liu2014}.   However, we must stress, as we found, that this is not always true. A close inspection shows that at weak confinement the velocity may  first  decrease and then
 increase; see below. {\color{black} It must be noted that besides swimming based on cilium or flagellum activity, there has been also a large number of studies devoted to amoeboid swimming bounded by walls in the biological literature, as discussed below}.

{\color{black} {Felderhof \cite{Felderhof2010} has reported on the fact that
 confinement enhances the speed of the Taylor swimmer. Later, Zhu {\it et al.} \cite{Zhu2013} considered the squirmer model to   show that speed  {\it decreases} with confinement when the squirmer surface deformation is tangential only. If on the contrary normal deformation is also allowed, the speed increases with confinement. Liu {\it et al.} \cite{Liu2014} reported on another model, namely a helical flagellum moving in a tube and  found that, except for  small tube radii,
the swimming speed for a fixed helix rotation rate increases monotonically with confinement. Acemoglu {\it et al.}\cite{Acemoglu2014} adopted  a similar model but, besides the flagellum, their swimmer is endowed with
 a head. They found in this case an opposite tendency: the speed {\it decreases}  with confinement. Bilbao {\it et al.}\cite{Bilbao2013} analyzed a model inspired by nematode locomotion by simulation and found that walls enhance the speed. Ledesma {\it et al.}\cite{Ledesma2013} treated a dipolar swimmer bounded both by rigid or an elastic tube and obtained  a speed enhancement due to walls.

 The effect of confinement on cell motility is a also a major field of research in biological literature. Several in vitro devices have been set up in order to analyze this issue for different kinds of cells. For example, some cancer cells use a mesenchymal motion when unconfined, while under confinement (and in the absence of focal adhesion) they can switch to a fast amoeboid  locomotion \cite{Liu2015}. The enhancement of the speed under confinement in this case is likely to be a complex phenomenon, since it is not only due to a pure hydrodynamical phenomenon, but to an internal cellular transition of mode locomotion. It will be interesting to include in our model  in the  future inclusion  several competing modes to check if one mode prevails over another when changing the external environment.} } 
  speed, pointing to the complexity of the effect of the walls.

\subsection{The flow field}
In this section we would like to provide a qualitative picture regarding the increase of swimming velocity with increased confinement.
A starting point for the qualitative argument describes the swimmer as grasping the wall to enhance its motion. When the confinement is too strong, the swimmer can not fully deploy
its shape and therefore a collapse of the velocity is expected. In actuality, the situation is more subtle, as will be discussed below.

Let us analyze in some details { the flow pattern} and dissipation around the swimmer to provide some hints toward understanding the basic mechanisms. We shall then point out some analytical results that show the complex nature of the phenomenon.
Fig. \ref{fig5} shows the flow pattern around the swimmer. It consists of two swirls which lie inside the AS.
 The flow pattern has a structure of a source dipole. A source in 2D creates a flow which is inversely proportional to distance in 2D, meaning that the source dipole has an asymptotic behavior which is inversely proportional to the square of the distance.


\begin{figure}[ptb]
\begin{center}
\includegraphics[width=0.9\linewidth]{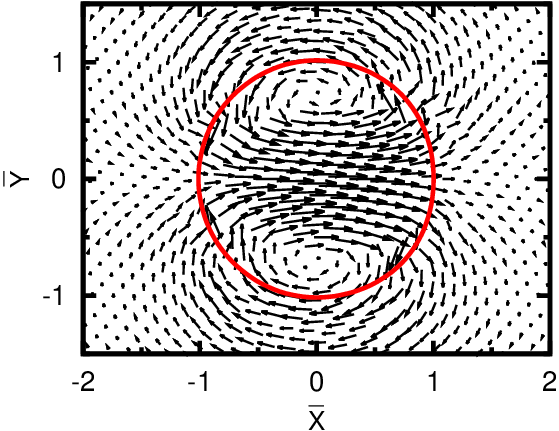}
\end{center}
\caption{\label{fig5}  (Color online) The time-averaged shape of the swimmer over one swimming cycle and
 its corresponding velocity field. $C=0.40$, $\Gamma=0.085$, $S=10$.  }%
\end{figure}


\subsection{Dissipation function}
\label{dissip}

\begin{figure}[ptb]
\begin{center}
\includegraphics[width=1.0\linewidth]{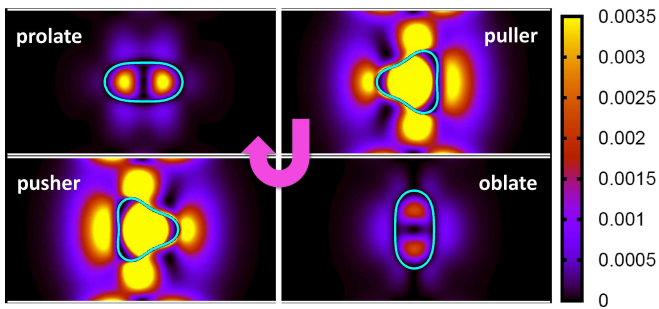}
\end{center}
\caption{\label{fig6} (Color online) A series of instantaneous dissipation
distributions   during one complete swimming
stroke. Color shows the intensity of dissipation density, and the blue represents the contour of the
swimmer. Swimming direction goes to the right. An amoeboid swimmer
combines  pusher and puller behavior during one deformation period. $C=0.40, \Gamma=0.085$, $S=10$. }%
\end{figure}

\quad The instantaneous intensity of dissipation density is given by:
\begin{eqnarray}
\Psi_v = \sigma_{ij} \frac{\partial u_i}{\partial x_j} = 2\eta \epsilon_{ij} \epsilon_{ij},
\end{eqnarray}
where $\sigma_{ij}=-p\delta_{ij}+2\eta\epsilon_{ij}$, $\epsilon_{ij}=1/2(\partial_j u_i+\partial_i u_j)$, and repeated subscripts are to be summed over. The expression reads explicitly
 in $x-y$ coordinates as
\begin{eqnarray}
\Psi_v = 2\eta \left[\left(\frac{\partial u_x}{\partial x}\right)^2+\left(\frac{\partial u_y}{\partial y}\right)^2\right]+\eta \left( \frac{\partial u_x}{\partial y}+\frac{\partial u_y}{\partial x}\right)^2.
\end{eqnarray}

The rate of work (output power) performed by AS is instantaneously  equal to the rate of total energy dissipation in the fluid.
The hydrodynamic interactions between the swimmer and the
walls are illustrated by a series of instantaneous dissipation
patterns produced during the process of one complete AS
stroke in Fig.~\ref{fig6}. The corresponding instantaneous configurations of
a wall-confined amoeboid swimmer are plotted by a closed cyan solid line (see Fig.~\ref{fig6}),
from which we can clearly observe that the swimmer uses the wall as a support on
which much of  the work is performed (grasping). The instantaneous dissipation shows two-arm-like distributions extended
to the walls during the puller and pusher states, so that one can intuitively understand that the
swimmer's speed is increased by extended arm-like dissipation structures creating hydrodynamical friction with the walls.

\begin{figure}[ptb]
\begin{center}
\includegraphics[width=0.9\linewidth]{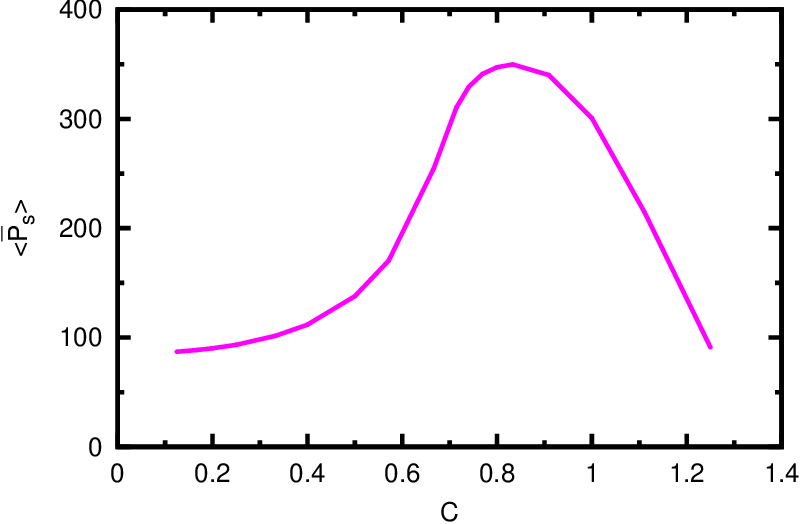}
\end{center}
\caption{\label{fig7} (Color online) The time-averaged power consumption  $\langle {\bar P_s}\rangle$  as a function of $C$ for $\Gamma=0.085$, $S=10$.}
\end{figure}

{\color{black} We define a time-averaged power consumption (rate of work) as
\begin{eqnarray}
   \langle P_s\rangle &=& \frac{1}{T_s}\int^{T_s}_{0} \oint_{\partial \Omega} \left[{\mathbf F}(s)\cdot {\mathbf u}(s)\right] d s d t \nonumber \\
   &=& \langle \oint_{\partial \Omega} \left[{\mathbf F}(s)\cdot {\mathbf u}(s)\right] d s \rangle,
\end{eqnarray}
\noindent{where $T_s$ is a swimming period.} A dimensionless average power is defined as $\langle {\bar P_s}\rangle=\langle P_s\rangle T_s^2/(\eta R_0^2)$.
Fig.~\ref{fig7}  shows the dimensionless time-averaged power of the active forces. Power consumption $\langle {\bar P_s}\rangle$} attains a maximum at a
confinement which is close (but not equal) to the confinement at which a maximum velocity is obtained (Fig.~\ref{fig3}).
%
In the strongly confined regime, the power
rapidly decreases because
 the deformation amplitude is  strongly reduced by the wall constraints: the swimmer is in principle able to deploy a larger amplitude but the rigid walls block this tendency.  This is supported by a recent experiment~\cite{Nosrati2015} according to which
  the flagellar wave amplitude of bull sperm
in the extremely confined condition is so strongly suppressed that they have to adopt a slithering motion to swim
along a rigid surface. The swimming velocity of bull sperm becomes slower and slower.


Note finally that an analytical study (a detailed report of which we make elsewhere\cite{Farutin2016}) performed at weak confinement has allowed us to extract the swimming velocity as a function of confinement:

\begin{equation}\label{vfinal}\langle {V}\rangle=\frac{5\pi\Gamma R_0}{3\sqrt{6}T_s}-0.12 C^2\frac{\pi\Gamma R_0}{T_s}+O(C^3\Gamma)+O(\Gamma^{3/2}).\end{equation}
The first term corresponds to the purely unconfined case and we capture the same dependence with $\Gamma$ as in the 3D case \cite{Farutin2013}. The second term is the first contribution of confinement to the swimming speed, and is negative, meaning that at very small confinement the wall reduces the swimming speed.
This is visible
in the inset of Fig. \ref{fig3}.
 This remark shows clearly that the effect of the walls is not quite obvious.

\section{Efficiency of swimming}
\label{effic}
In this section we wish to investigate the notion of efficiency.
We will first adopt a widely used definition of
swimming efficiency, which is the ratio of the least power required to drag (or pull) the swimmer along
the axis at its time-averaged speed $\langle V\rangle$ (the physical one, not the dimensionless one) over the actual time-averaged output power $\langle P_s \rangle$
generated by the swimmer~\cite{Lighthill1975,Purcell1997,Becker2003}.
Here we define a dimensionless $2$D efficiency  $\Pi_1$ as
\begin{equation}
\Pi_1 = \frac{\eta \langle V\rangle^2}{\langle P_s \rangle},
\label{effi1}
\end{equation}
where $\langle V\rangle$
is the average   physical velocity.
¨
Figure~\ref{fig9} shows that the optimal swimming occurs at the transition point where  the velocity as a function of the applied force saturates (see Fig.\ref{fig8}). The qualitative behavior of the efficiency curve can be explained
as follows. When $S$ is small (small force, before the plateau is reached in Fig.\ref{fig8}) the stroke dynamics is so fast that it does not allow to the swimmer
to attain its saturated shape, and the swimmer loses efficiency. If $S$ is too large, this does not bring any benefit since the swimmer  reaches its saturated shape too quickly, and does not gain in efficiency by a further increase of force.

\begin{figure}[ptb]
\begin{center}
\includegraphics[width=0.9\linewidth]{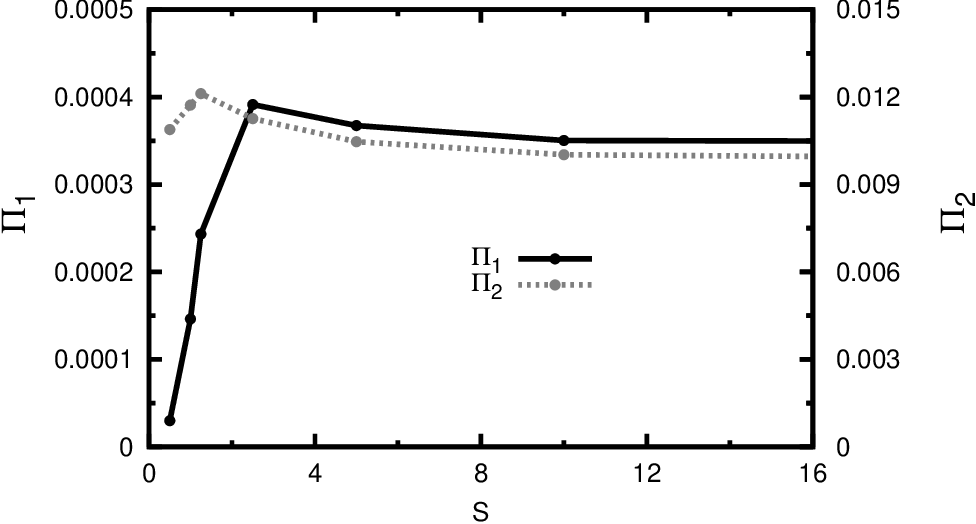}
\end{center}
\caption{\label{fig9} (Color online) The two efficiencies  $\Pi_1$ and $\Pi_2$ as functions of $S$.  $C=0.5$ and $\Gamma=0.085$.}%
\end{figure}

\begin{figure}[ptb]
\begin{center}
\includegraphics[width=0.9\linewidth]{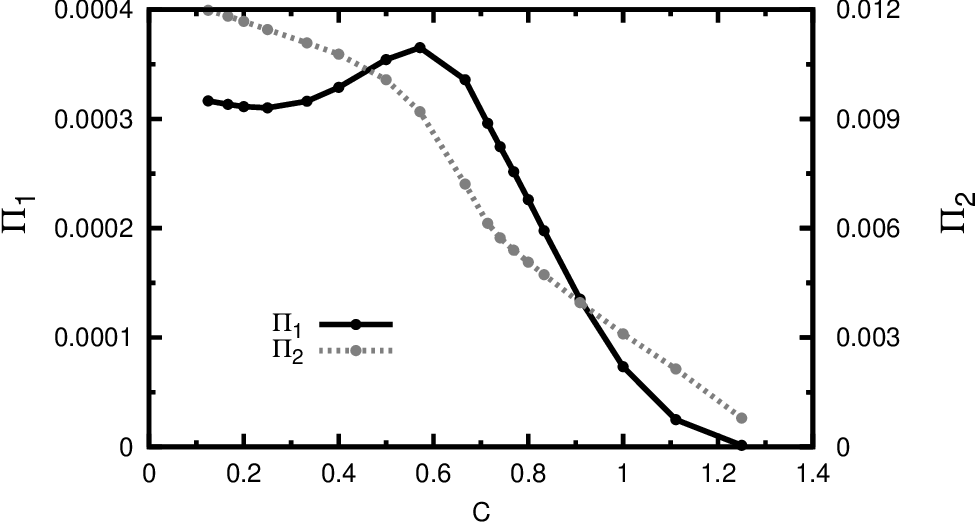}
\end{center}
\caption{\label{fig10} The efficiencies  ${\Pi}_{1}$ and $\Pi_2$   as  functions of $C$ for $\Gamma=0.085$, $S=10$.}
\end{figure}

It is also worthwhile to examine the efficiency as a function of the strength of confinement $C$. Using the definition above, we find that the maximum efficiency
occurs at a confinement of $C\simeq 0.6$, which is different from
 that corresponding to the maximum speed (Fig. \ref{fig3},  $C\simeq 0.75$).
The efficiency ${\Pi}_{1}$ first gradually increases and then decreases rapidly
after the optimal point.

{ A second swimming efficiency considered in literature is noteworthy. Defined as $\sim \langle V\rangle/\langle P_s \rangle$~\cite{Shapere1987,Koiller1998}, it seems to be suitable for small deformations since for small deformations, $V\sim \Gamma$, then $P_s\sim \Gamma$; the efficiency is independent of $\Gamma$. This definition of efficiency is not dimensionless.
We adopt here a dimensionless form given by
\begin{equation}
\Pi_{2} = \frac{\eta\omega R_0 \langle V\rangle}{\langle P_s \rangle}.
\label{effi2}
\end{equation}

The corresponding results are shown by the grey dashed dotted line in Fig.\ref{fig7} and Fig.~\ref{fig10}, from which
no optimal value for a special  confinement is found, in contrast to the definition (\ref{effi1}).  This example clearly highlights the main difference between the two definitions. It also points to the fact that there seems to be no straightforward definition of a suitable efficiency (if any).

\section{Various dynamical states of the swimmer: from straight trajectory to navigation}
\label{various}

In this section we present the rich panel of behaviors manifested by the AS.
We shall see that the AS can settle into a straight trajectory, can navigate in the
channel with a fixed amplitude in a symmetric or asymmetric way, or can even crash
into the wall. The adoption of one or another regime depends on various conditions, and especially on the force
distribution and on the degree of confinement. One extra degree of complexity that
is unique to the AS is the fact that the nature of the swimmer (puller, or pusher) is
not an intrinsic property of the swimmer itself, but depends  on the environment
(say on confinement). Before discussing our results, we would like first to put our work in the context of swimmers
in general.

Zhu {\it et al.} \cite{Zhu2013} adopted
the squirmer model which can be set to be a pusher, a puller or neutral, by the appropriate choice of parameters. They find
 that a pusher crashes into the wall, a puller settles into a
straight trajectory, and a neutral swimmer navigates. However, the navigation amplitude depends on initial conditions:  it is not a limit cycle, but an oscillator, akin to what happens for Hamiltonian systems.

Najafi {\it et al.} \cite{Najafi2013} used a three-bead model and reported on the navigation possibilities of the swimmer. {It was not clear in that study
if the navigation amplitude was dependant or not on initial conditions. Actually, no conclusion on whether or not the navigation period reaches a final given value was made.

Since the nature of the swimmer seems to play an important role,
it is interesting to first define  this notion and analyze it  for our system before presenting the main results.

\subsection{Swimmer nature evolution}

\begin{figure}[ptb]
\begin{center}
\includegraphics[width=0.9\linewidth]{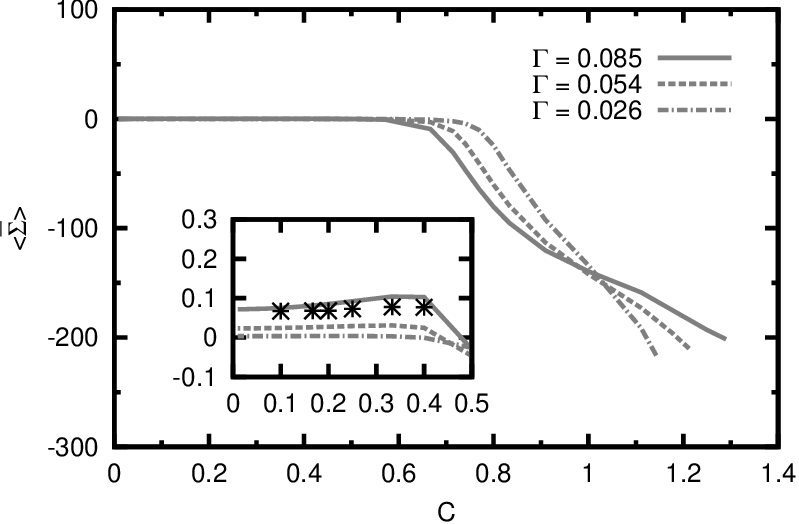}
\end{center}
\caption{\label{fig16} (Color online) Time-averaged stresslet $\langle\bar \Sigma\rangle$ as a function
of confinement $C$ showing the transition from pusher to puller.  Different lines refer to  an axially moving swimmer
and stars  to a symmetrically navigating swimmer. Inset shows a zoom for small $C$ in order to show the confinement at which the pusher-puller transition occurs. $S=10$.}%
\end{figure}

{\color{black} Swimmers are classified as pushers, neutral swimmers or pullers, depending on the sign of the stresslet. This notion is defined on the average over one cycle. During half a cycle a swimmer may behave as a puller while during the second half as a pusher. What matters here is the average nature. For example, {\it Chlamydomonas Reinhardtii}~\cite{Drescher2009,Guasto2010,Garcia2011,Kantsler2013}, switches from a puller to a pusher over one cycle, but
  is a puller on  the average over one swimming cycle. The nature of our swimmer will also be defined by an average over one cycle.
The interesting feature which emerges here is that our swimmer may behave as a pusher on the average, but if the confinement changes it switches, on average, to puller behavior. In other words, the swimmer seems to adapt its nature to the environment.}
}
In the far-field approximation, the first contribution (in a multipolar expansion) to the velocity field is governed by a term of the form  $\Sigma_{ij}=\oint F_ir_j ds$. Exploiting symmetry of the swimmer (axial symmetry) we can write that the far field is, to leading order, proportional to
$\bar\Sigma=(\Sigma_{xx}-\Sigma_{yy})/(\eta R_0^2/T_s)$, which we will call a (dimensionless) stresslet hereafter.
The instantaneous type (pusher or puller) of an AS can be identified by the sign of the instantaneous stresslet:
(${\bar\Sigma}>0$ indicates a pusher and ${\bar\Sigma}<0$ indicates a puller). As already pointed out earlier \cite{Farutin2013} our swimmer observes, in the course of time, an entangled puller-pusher state. We could split the swimming cycles into 4 intervals. In the first interval the swimmer behaves  as a pusher, followed by an interval where it behaves as a mixed pusher-puller state, a third interval in which it behaves as a puller, and finally a fourth interval in which it assumes a mixed state (see also Fig.\ref{fig6}). Another possible way of characterizing our swimmer is to determine its nature on average over one swimming cycle. {\color{black} The dimensionless average stresslet is} given by ${\langle \bar\Sigma \rangle}=(\int _0^{T_s} \bar\Sigma dt)/T_s$ (the average over
one deformation period $T_s$). Figure \ref{fig16} shows this quantity as a function of confinement.
The swimmer behaves as a pusher for  $C\lesssim 0.5$, and as a puller for $C>0.5$. The efficiency shown in Fig. \ref{fig10} shows an optimum for $C\sim 0.6$ which is close to the pusher-puller transition.


\subsection{Navigation and symmetry-breaking bifurcation}

In the weak confinement regime (where the swimmer is a pusher on  average, see Fig. \ref{fig16}) it is found that the straight trajectory is {unstable in favor of  navigation}, and that the navigation amplitude is a function of confinement independent of initial conditions. This is confirmed by the results of both numerical methods (BIM and IBM). Snapshots  are shown in Fig. \ref{fig11} where one can observe ample navigation of the swimmer. Typical evolution of the $Y_c$ position as a function of time is shown in Fig. \ref{fig11}. This curve shows a large scale navigation period with a small scale structure associated with the strokes of the swimmers.
Let us define the navigation amplitude $\Delta Y$ as the difference between the maximum of $Y(t)$, $Y_t$, and the minimum of $Y(t)$, $Y_b$, $\Delta Y= Y_t-Y_b$ and plot this amplitude  as a function of confinement $C$ (Fig. \ref{amplitude_C}). We see that the amplitude decreases with $C$. The swimmer does not crash into the wall, in contrast with the squirmer model studied in Ref. \cite{Zhu2013} . Indeed, we find that the navigation amplitude reaches saturation after some time.
The  navigation mode also survives when the swimmer behaves as a puller on  average, and does not settle into a straight trajectory, unlike the finding in Ref. \cite{Zhu2013}. In  Fig.  \ref{amplitude_C}  the domains of puller and pusher are shown.

The navigation mode undergoes an instability at a critical confinement $C\simeq 0.8$ (where it behaves as a puller)  in favor of an asymmetric motion where the swimmer moves closer to one of the two walls. The amplitude $Y_c(t)$ of the center of mass  is shown in Fig. \ref{symm_bifurcation}. The center of mass shows an oscillation which is determined by the stroke frequency, in contrast to in the navigating mode which, apart from  the  small scale oscillation due to the strokes, develops large scale oscillations of the position. {\color{black} Figure} \ref{symm_bifurcation} shows the trajectory of the asymmetric swimmer. The overall behavior
is summarized in Fig. \ref{fig17}. There we show the average  position of the center of mass
 as a function of confinement $C$. For small $C$, the center of mass position averaged over a navigation period  is  zero.  There
 is   a spontaneous  symmetry-breaking bifurcation at $C\sim 0.8$ where the swimmer moves towards either of the two walls. In this regime the center of mass of the swimmer shows temporal oscillations which are asymmetric (lower rectangle in the figure). The bifurcation is supercritical (albeit quite abrupt).
 At  $C =1.04$, the average position of the center of mass  crosses zero.

\begin{figure}[ptb]
\begin{center}
\includegraphics[width=0.9\linewidth]{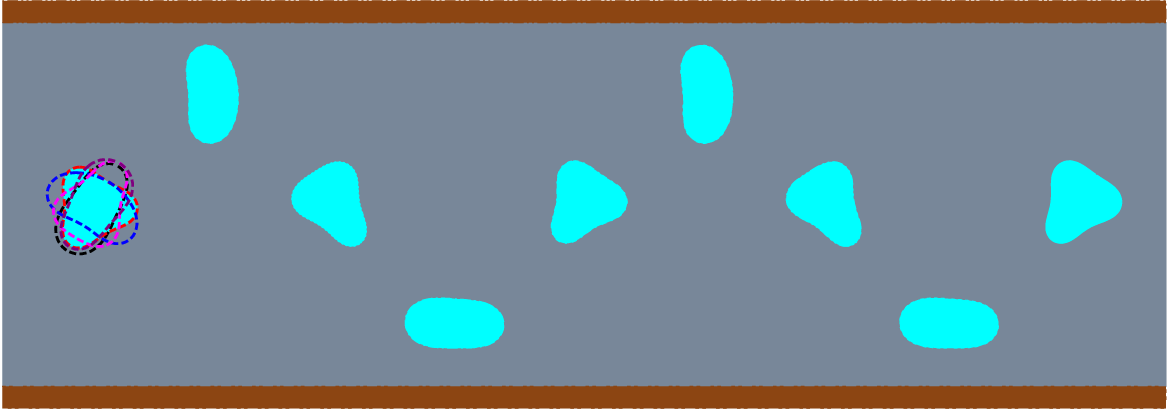}
\includegraphics[width=0.9\linewidth]{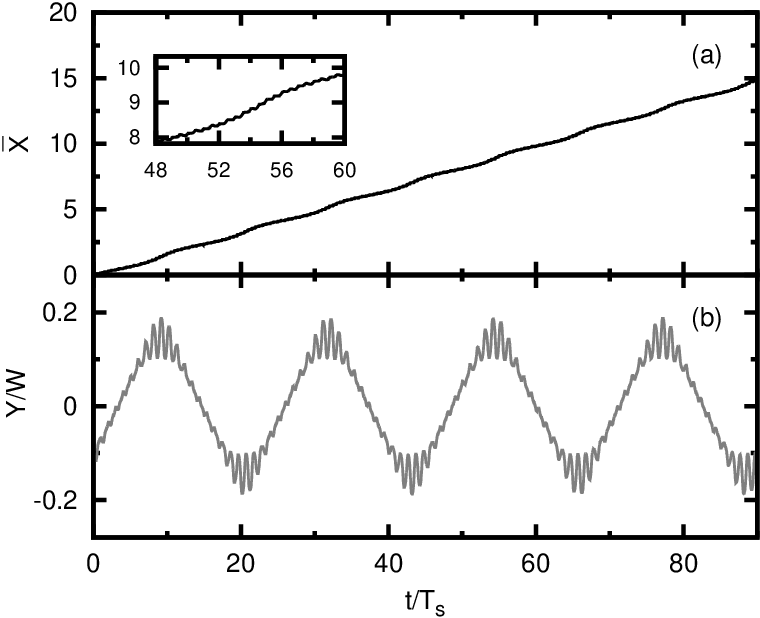}
\end{center}
\caption{\label{fig11} (Color online) Top: Snapshots of a navigating swimmer over time ($W=10.0R_{0}$).
The dashed profiles show a complete period $T_s$ of deformation. Few shapes are
represented over two navigation periods $T$ of the order of $240 T_s $. $\Gamma=0.085$, $S=10$. %
 Bottom: A navigating amoeboid swimmer in the channel
undergoes  periodic oscillations in both ${x}$ (a) and ${y}$ (b)
directions. The inset in (a) shows the short periodic oscillations
due to strokes which are superimposed on the long periodic oscillations due to the
navigation.   $C=0.50, \Gamma=0.085$, $S=10$. }%
\end{figure}


\begin{figure}[ptb]
\begin{center}
\includegraphics[width=0.9\linewidth]{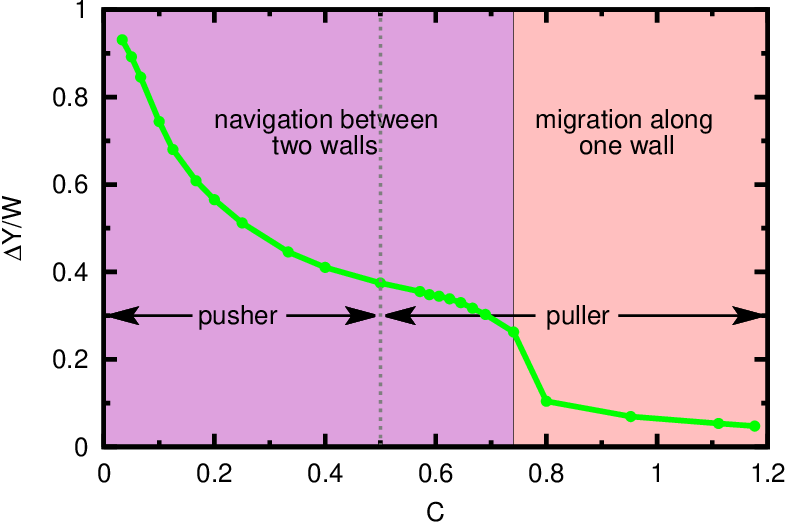}
\end{center}
\caption{\label{amplitude_C} (Color online) The amplitude of navigation as a function of confinement $C$. $\Gamma=0.085$, $S=10$.}%
\end{figure}


\begin{figure}[ptb]
\begin{center}
\includegraphics[width=0.9\linewidth]{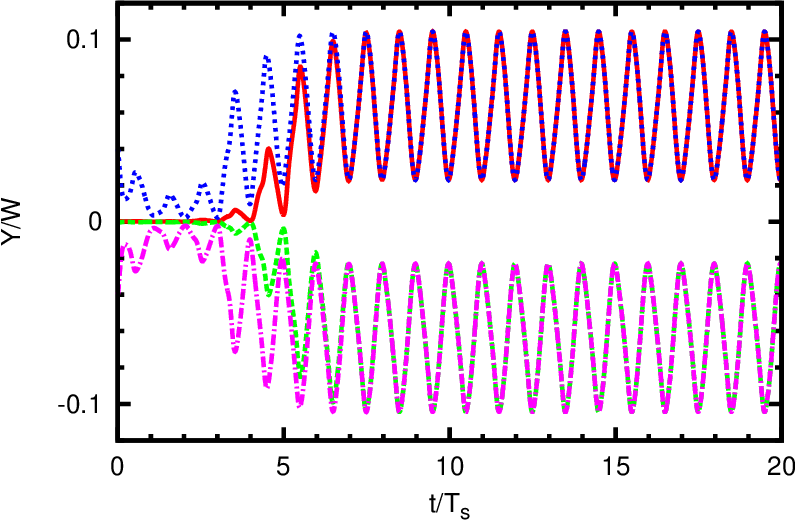}
\includegraphics[width=0.9\linewidth]{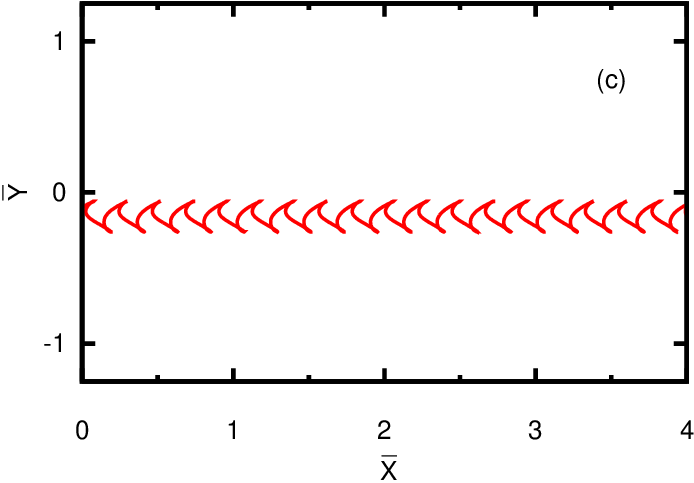}
\end{center}
\caption{\label{symm_bifurcation} (Color online) Top: The vertical position of center of mass $Y_c$ as a function of time $t$ showing the symmetry breaking bifurcation.
Different colors represent different initial positions. Bottom: trajectory of the asymmetric swimmer. $\Gamma=0.085$, $S=10$, $C=0.8$.}%
\end{figure}

\subsection{Ability of scanning space}

We were interested in analyzing the swimmer's ability to scan the available space. We measured the highest  membrane point of the swimmer in the channel, calling it $H_t$ and the lowest membrane point, calling it $H_b$. The difference
$\Delta H= H_t-H_b$ is a direct measure of space scanning by the swimmer. In Fig.~\ref{fig18}  (top panel) we provide a schematic representation of this measure. The lower panel provides the numerically computed scanning amplitude as a function of confinement.
 It is found that the navigating mode can
explore between about $80\%$ and  $95\%$ of the available lateral space.

\subsection{Crashing into the wall and settling into a straight trajectory}

As stated above it has been reported \cite{Zhu2013} (by adopting a squirmer model) that the pusher crashes into the wall while a puller settles into a straight trajectory. This contrasts with our finding presented above. We have thus attempted to investigate this question further in order to clarify the situation. We find that the type of trajectory depends strongly on the strength of the stresslet {\color{black} $\bar\Sigma$ and the strength of a linear combination $D$ of dimensionless  force quadrupole $\bar\Sigma_2$ and source dipole $\bar\Delta_2$ \cite{Wu2015}.} A weak pusher or puller means that the stresslet $\bar\Sigma^2 \ll  \bar V \bar D $ \cite{Wu2015},  where $\bar D$ is the dimensionless force
quadrupole strength.  For example, if the swimmer  is a weak pusher, we have navigation and symmetry-breaking as presented above.} On the contrary, if the pusher is strong enough we also find that the swimmer has a tendency to crash into the wall. Similarly, if the puller is strong enough we find that the swimmer settles
into a straight trajectory with $Y_c(t)=0$ if the swimmer is at the center of the channel, or with $Y_c(t)$ periodic in time due to strokes if the swimmer at some distance from  the center, as is the case  in Fig. \ref{symm_bifurcation}, bottom.

\begin{figure*}[ptb]
\begin{center}
\includegraphics[width=0.9\linewidth]{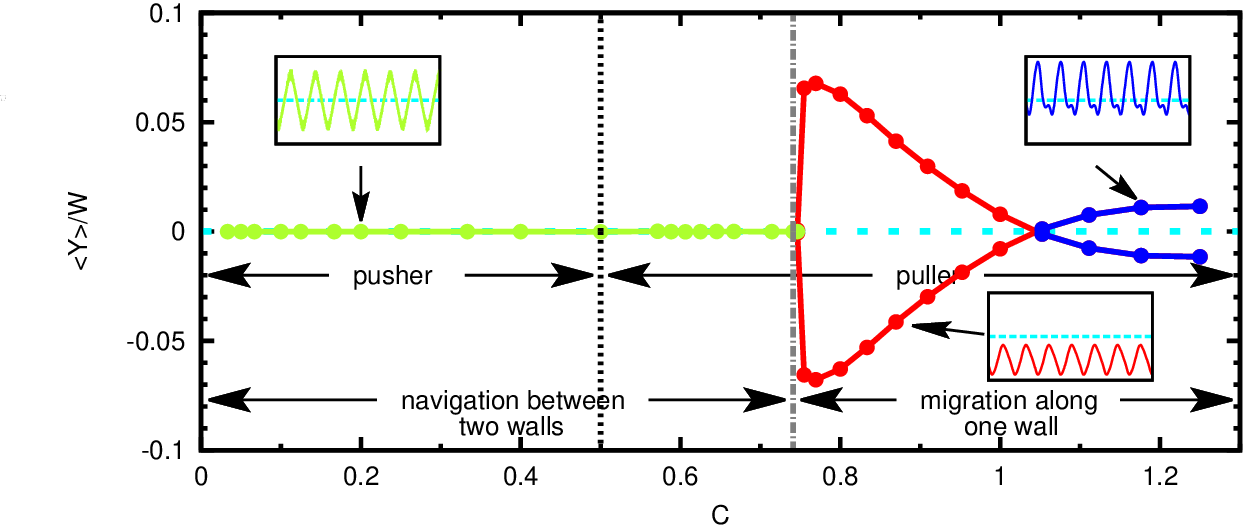}
\end{center}
\caption{\label{fig17} (Color online) {\color{black} Average position along $y$ as a function of confinement. $\Gamma=0.085$, $S=10$.}}
\end{figure*}

In order to have a swimmer with a large enough stresslet, we can modify the force presented in section \ref{model} in the following way
 \begin{equation} {\mathbf F} =2 [ \sin(\omega t) \cos (3\alpha)-(\beta + \cos (\omega t )) \cos(2 \alpha)] {\mathbf n}
 \label{puller}
  \end{equation}
   where we recall that $\alpha =2\pi s/L_0$ ($s$ is the arclength and $L_0$ the perimeter), or

  \begin{equation} {\mathbf F }=2[ (-\beta + \sin(\omega t)) \cos (3\alpha)- \cos (\omega t ) \cos(2 \alpha)] {\mathbf n}.\label{pusher}
    \end{equation} Setting $\beta =0.5$ in equation (\ref{puller}) and (\ref{pusher}) gives a puller and a pusher, respectively.

Fig. \ref{crash} shows a typical trajectory for the case where the strong pusher crashes into the wall.  We also provide the corresponding values of the stresslet. It would be interesting in the future to make a more detailed analysis of this problem in order to determine the  critical amplitude of the stresslet above which we have a transition from navigation to crashing/settling into a straight trajectory. {\color{black} In contrast, a strong puller is found (Fig. \ref{symm_bifurcation}, bottom) to move along the channel. The center of mass undergoes small vertical oscillation due to shape deformation when the swimmer selects an off-centered  trajectory.}

\section{Discussion and perspectives}
\label{conclusion}

The current work presents a systematic study of the hydrodynamics of microswimming
by shape  deformation (amoeboid swimming) in a planar microfluidic channel. By studying the
coupling of shape change and wall interaction, we discovered several new features of microswimming.
{\color{black} Let us highlight the new features discovered here, and recall some facts we reported on recently}.
\begin{itemize}
\item {\color{black} In agreement with our recent report \cite{Wu2015} we confirm that the nature (pusher or puller) } of the swimmer is not an intrinsic property. Several swimmers may have an instantaneous nature which changes over time. For example, {\it Chlamydomonas Reinhardtii} ~\cite{Drescher2009,Guasto2010,Garcia2011,Kantsler2013} is on the average (average over one swimming cycle) a puller, albeit it has two temporal phases where it behaves like a puller or a pusher.  We also found here  that the amoeboid swimmer can exhibit a pusher or a puller behavior at a given time; however, the stresslet, averaged over one cycle, can change under the influence of external factors such as confinement, unlike other swimmers studied so far. For example, by changing the environment (like confinement) an amoeboid swimmer which is a pusher on the average, becomes a puller (or vice versa) when external parameters are changed.

\item The straight trajectory of the swimmer can become unstable in favor of navigation. Navigation has been reported for other swimmers in two different studies \cite{Najafi2013,Zhu2013}. However, in the  study using a three-bead model \cite{Najafi2013} there is no conclusion about the final amplitude of navigation (it increases monotonously with time), while in the  study using a squirmer as a model \cite{Zhu2013} navigation was found for a neutral swimmer only, and the navigation amplitude depended on initial conditions. Here, in contrast, we found that navigation can occur for pushers (a weak enough pusher) as well as for neutral swimmers. For pushers we find that the amplitude is fixed, in that  it does not depend on initial conditions.

    \item The confinement was found to enhance the swimming speed in several previous studies \cite{Felderhof2010,Jana2012,Zhu2013,Bilbao2013,Ledesma2013,Acemoglu2014,Liu2014}. Here also we reach the same conclusion, but with the specification that monotonous behavior (even in the weak confinement regime) is not a general tendency, so the reverse can also sometimes occur (Fig. \ref{fig3}). This points to the fact that this question should be more carefully considered.

            \item In a previous study \cite{Zhu2013} it was reported that a pusher crashes into the wall and a puller settles into a straight trajectory. Here we have reported that this is not a general result: a pusher can also undergo navigation with a well-defined amplitude.
                \item The swimming speed is found to  be quadratic with the force amplitude and then tends towards a plateau at large enough amplitude. The quadratic behavior is shared with that of a three-bead swimmer \cite{Pande2015}, but no  regime with a plateau was reported there. We believe that a modified three-bead model having a nonlinear spring with saturation (like with the so-called FENE model) should capture the plateau behavior.
\end{itemize}}
{\color{black}
Although the present study provides a first route for the modeling of amoeboid swimming, several other questions deserve future consideration in order to capture a more realistic picture of swimming cells.
In the present work, we assumed the membrane to be a simple phospholipid envelope whereas real cells are endowed with a cytoskeleton and a nucleus making the deformations more difficult.
From equation (\ref{scalinsmallf}), one can extract an order of magnitude of the active force of about $0.01$ pN (using typical quantities \cite{Liu2015} for a $10 \mu$m cell diameter, a migration velocity of $10 \mu$m/min, a water-like medium viscosity and a swimming cycle of characteristic time $1$ min). This force is smaller than the available force in a real cell. It will be of great importance to include the cell cytoskeleton in modeling for better comparison with experimental systems.
For the sake of simplicity we have also considered a simple periodic function for the active forces. Real observations of cell motion \cite{Lammermann2009,Liu2015} point to more complex shape dynamics. Similar complex motions could be reproduced by assigning a complex time-dependent force distribution (including higher harmonics). It will be interesting in the future to use experimental data and, by solving an inverse problem, extract the active forces (both in space and time) to be used in our model for a more realistic analysis.}

 Another future issue is the study  of the wall compliance effect \cite{Ledesma2013}. Indeed, many eukaryotic cells {\it in vivo} (e.g. cells of the immune system, cancerous cells...) move in soft tissues where cells are often in interstices, in between other cells and extracellular matrix. Therefore, the confinement is not fixed in time but varies in response to stresses created by the cells. A basic representation of this effect could be had by considering bounding walls of a certain compliance. Flexible walls can promote swimming  in the strong confinement regime by allowing larger amplitude of deformation of the swimmer than rigid walls do.
 Another important issue is to study the collective behaviors. For example, during wound healing and cancer spreading (to name but a few examples) cells move in a concerted fashion. It would be thus important to analyze if collective behaviors could emerge from the study of multiple swimmers, such as special patterns, synchronization, and so on. Finally,  {\it in vivo} cells move in complex visco-elastic materials, and it will be an important issue in the future to analyze AS in a complex medium. {\color{black} An interesting issue will be to analyze the motion of a cell surrounded by other cells representing tissues. For example, how  the rheology or the visco-elatsic properties of the tissue affect the amoeboid motion, both for a single cell and for collective motions.}

\begin{figure}[ptb]
\begin{center}
\includegraphics[width=0.9\linewidth]{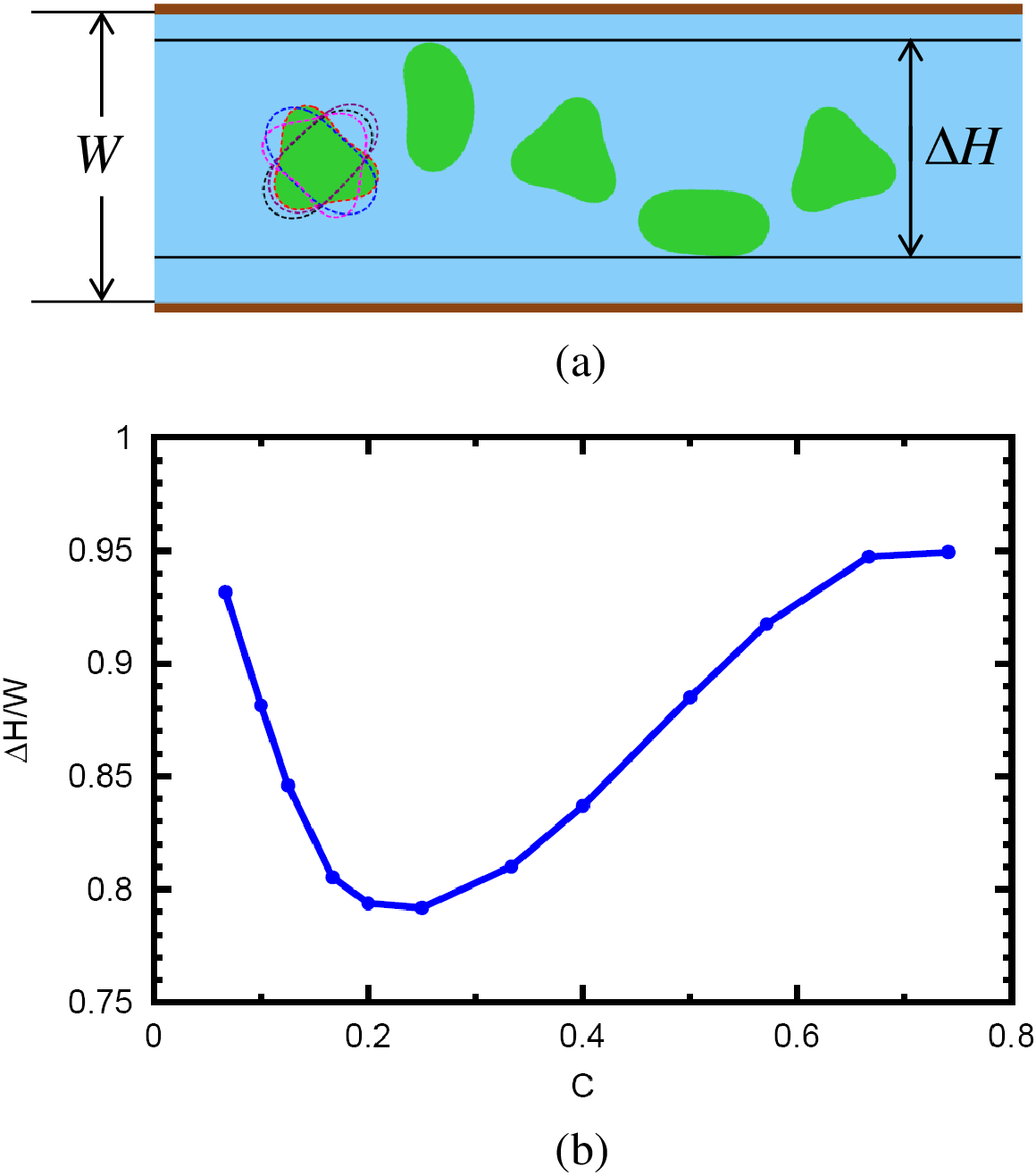}
\caption{\label{fig18} (Color online) (a) The  definition of height difference $\Delta H$
scanned by a navigating swimmer. (b) The scanned space  for different confinements ($\Gamma=0.085$, $S=10$).}
\end{center}
\end{figure}

\begin{figure}[ptb]
\begin{center}
\includegraphics[width=0.9\linewidth]{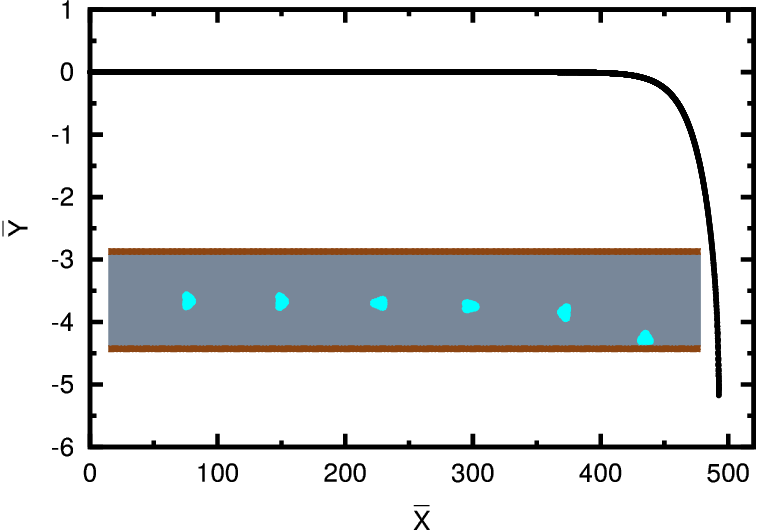}
\end{center}
\caption{\label{crash} (Color online) A trajectory showing a swimmer crashing into the wall. $\Gamma=0.054$, $S=1$, $C=0.167$.}
\end{figure}

%
%

\section{Acknowledgments}
H.W. thank Prof. S. Jung (Virginia Tech) for the helpful discussion on the experimental details.
C.M., A.F., M.T., and H.W. were supported by CNES and ESA. S.R. and P.P. acknowledge support from ANR.
All the authors acknowledge the French-Taiwanese ORCHID cooperation grant. W.-F. H.,
M.-C.L., and C.M. thank the MoST for a support allowing initiation of this project.

\section*{Appendix : Total force and total torque of tension force are zero} 
We have seen from general considerations that if the energy does not depend on position in space, then the total associated forces and torque are automatically zero. For interested readers we briefly give here an explicit proof.
In $2$D space, an amoeboid swimmer is represented by a $1$D closed contour. For the force this is quite obvious since that force can also be written as
 $$\mathbf{F}^{\rm{ten}}=\frac{\partial(\zeta \mathbf{t})}{\partial s}$$
 Therefore the integral over a contour is zero.


For the total  torque $\mathbf{T}^{\rm{tot}}$ along the perimeter of amoeboid swimmer, we have:
 \begin{eqnarray*}
 \mathbf{T}^{\rm{tot}} &=& \oint_{\partial \Omega} \left[\mathbf{r}\times\mathbf{F}^{\rm{ten}}\right] \rm{d} s  \nonumber \\
 &=& \oint_{\partial \Omega} \left[(x-X_c)F^{\rm{ten}}_y- (y-Y_c)F^{\rm{ten}}_x\right]
 \rm{d} s \\
 &=& \oint_{\partial \Omega} \left[(x-X_c)\frac{\partial(\zeta t_y)}{\partial s}- (y-Y_c)\frac{\partial(\zeta t_x)}{\partial s}\right]
 \rm{d} s \\
 &=& \oint_{\partial \Omega} \left[(x-X_c)d(\zeta t_y)- (y-Y_c)d(\zeta t_x)\right] \\
 &=& \oint_{\partial \Omega} \left[-\frac{\partial(x-X_c)}{\partial s}(\zeta t_y)+ \frac{\partial(y-Y_c)}{\partial s}(\zeta t_x)\right] \rm{d} s\\
 &=& \oint_{\partial \Omega} [-t_x(\zeta t_y)+ t_y(\zeta t_x)] \rm{d} s \\
 &=& 0
 \end{eqnarray*}

\footnotesize{
\bibliography{amoeba}
}

\end{document}